\documentclass[twocolumn,showpacs,preprintnumbers,amsmath,amssymb]{revtex4-1}
\usepackage{comment} 
\usepackage{graphicx}
\usepackage{dcolumn}
\usepackage{bm}
\usepackage[compact]{titlesec}
\usepackage{float}
\usepackage{subfig}
\usepackage{lipsum}

\usepackage{xcolor}

\usepackage{hyperref}
\newcommand{\rmd}{{\rm d}}

\begin{document}

\preprint{CTPU-PTC-23-24}
\title{Hairy Black Holes by Spontaneous Symmetry Breaking}

\author{Boris Latosh}
\email{latosh.boris@ibs.re.kr}

\author{Miok Park}
\email{miokpark76@ibs.re.kr}
\affiliation{Particle Theory  and Cosmology Group, Center for Theoretical Physics of the Universe, Institute for Basic Science (IBS), Daejeon, 34126, Korea}

\begin{abstract}
  \noindent
We study hairy black hole solutions in Einstein(--Maxwell)--scalar--Gauss--Bonnet theory. The scalar coupling function includes quadratic and quartic terms, so the gravitational action has a U(1) symmetry. We argued that when the effective mass of the scalar field is at the critical value, the non-hairy black holes transform into hairy black holes in a symmetry-broken vacuum via spontaneous symmetry breaking. These hairy black holes are stable under scalar perturbations, and the Goldstone bosons are trivial. Moreover, we found that the spontaneous symmetry breaking associated with local U(1) is unlikely to occur in this theory.  
\end{abstract}

                             
\maketitle

\section{Introduction}

The detection of gravitational waves from the merger of binary black holes by the Laser Interferometer Gravitational-Wave Observatory (LIGO) \cite{LIGOScientific:2016aoc} was a major breakthrough in recent decades. One of the missions of gravitational waves is to test general relativity since it alone struggles to explain the presence of dark matter, dark energy, and inflationary expansion \cite{Clowe:2006eq,Linde:1983gd,Starobinsky:1980te}. As an alternative to general relativity \cite{Lovelock:1971yv,Kobayashi:2011nu,Copeland:2012qf,Babichev:2022djd,Dorlis:2023qug}, we consider the Einstein(--Maxwell)--scalar--Gauss--Bonnet theory which has a non-minimal coupling of the scalar field with the Gauss-Bonnet(GB) term. The theory belongs to Horndeski gravity and has second-order field equations, so it is free of the ghost problem. Additionally, the evasion of the no-hair theorem was first studied in \cite{Antoniou:2017acq} and later in \cite{Lee:2018zym} based on Bekenstein's argument \cite{PhysRevLett.28.452,PhysRevD.51.R6608}. The complete derivation for the evasion of the no-hair theorem was done in \cite{Papageorgiou:2022umj}. At the same time as the discovery of hairy black holes in \cite{Antoniou:2017acq}, spontaneous scalarization was proposed to explain how hairy black holes acquire their scalar hair from the non-hairy ones \cite{PhysRevLett.120.131104}. This mechanism relies on a tachyonic instability that triggers the spontaneous growth of a scalar hair on a non-hairy black hole background  \cite{PhysRevLett.120.131104}. However, the produced hairy black holes are unstable under the perturbation of scalar fields \cite{Blazquez-Salcedo:2018jnn,Doneva:2017bvd}. Later studies showed that the coupling function with quadratic and quartic terms can generate stable hairy black holes in some parameter regimes \cite{Minamitsuji:2018xde, Macedo:2019sem, Minamitsuji:2023uyb}. Recently stable spontaneous scalarization for a quadratic coupling is suggested in \cite{Antoniou:2022agj, Kleihaus:2023zzs}.

We study a U(1)-invariant theory with a single complex scalar field and a non-minimal coupling to the GB term:
\begin{align}
  f(\varphi^*, \varphi) = \alpha \, \varphi^{*}(r) \varphi(r) - \lambda \, \big( \varphi^{*}(r) \varphi(r) \big)^2. \label{eq:f}
\end{align}
This allows us to study hairy black holes in symmetric and symmetry-broken phases. We define the \textit{symmetric phase} as the phase in which the scalar fields near the horizon are at either the ``global" minimum ($\alpha < 0$) or the ``local" maximum ($\alpha > 0$) of the interacting potential ($V=-f(\varphi^*, \varphi)\, \mathcal{G}$). The \textit{symmetry-broken phase} is the phase in which the scalar field near the horizon is at the ``global" minimum ($\alpha > 0$). In contrast to spontaneous scalarization, which requires a negative effective mass squared to generate hairy black holes, we show that stable hairy black holes are generated in the symmetry-broken phase when the effective mass squared is positive. Thus we provides a mechanism for generating stable hairy black holes rooted in the symmetry of the theory.

This paper is organized as follows. Section 2 discusses the global U(1) symmetric theory and shows that the Schwarzschild black hole becomes unstable beyond $\alpha_{\textrm{Sch.}}$ against the scalar field perturbation.
In Section 3, we find hairy black holes in symmetric and symmetry-broken phases and investigate their instability. We calculate the mass and scalar charge of those hairy black holes. In Section 4, we study electrically charged hairy black holes by spontaneous symmetry breaking in the theory with local U(1) symmetry. Section 5 summarizes our results.\\

\section{Generation of hairy black holes}

We consider the action in four-dimensional asymptotically flat spacetime:
\begin{align}
  & S = \int \textrm{d}^4 x \sqrt{-g} \bigg[ \frac{R}{2 \kappa^2}  - \nabla_{\alpha} \varphi^{*} \nabla^{\alpha} \varphi + f(\varphi^*, \varphi )\, \mathcal{G}  \bigg], \label{eq:action1}\\
  & \mathcal{L}_{\varphi} = - \nabla_{\alpha} \varphi^{*} \nabla^{\alpha} \varphi \!+\! f(\varphi^*, \varphi)\, \mathcal{G} = T \!-\! V, \\
  &V = - f(\varphi^*, \varphi)\, \mathcal{G} . \label{eq:action2}
\end{align}
Here $\mathcal{G}$ is the GB term
\begin{align}
   \mathcal{G} = R_{\mu \nu \rho \sigma} R^{\mu \nu \rho \sigma} - 4 R_{\mu \nu} R^{\mu \nu} +R^2 ,
\end{align}
the coupling function is given by (\ref{eq:f}), $\varphi(r)$ is a complex field, and $\alpha$, $\lambda$ are coupling constants. We consider $\lambda$ positive and allow $\alpha$ to take any real values. This Lagrangian respects the global $U(1)$ symmetry 
\begin{align}
\varphi(r) \rightarrow e^{i \chi} \varphi(r)
\end{align}
where $\chi$ is a constant. We use the metric ansatz 
\begin{align}
  \rmd s^2 = - A(r) \rmd t^2 + \frac{1}{B(r)} \rmd r^2 + r^2 \big(\rmd \theta^2 + \sin^2\!\theta\, \rmd \phi^2\big).
\end{align}
 
We consider a linear perturbation of the scalar field around the background solution to examine the instability. The perturbation equation is written
\begin{align}
  \bigg( \nabla_{\alpha} \nabla^{\alpha}  + f_{\varphi^* \varphi}\, \mathcal{G} \bigg) \delta \varphi (r)+ f_{\varphi^* \varphi^*} \, \mathcal{G} \, \delta \varphi^*(r)  = 0, \\
m_{\textrm{eff}}^2 = - \, f_{\varphi^* \varphi} \, \mathcal{G} ,
\end{align}
where the subscript of $f$ indicates a derivative with respect to corresponding variables. To simplify the analysis, we decompose the complex scalars into real scalars in (\ref{eq:CtoR}) and impose $\varphi_1 = \varphi_2$ for simplicity. We replace the perturbed field with the following substitution
\begin{align}
  \delta \varphi_1(t,r,\theta,\phi) = \sum_{l,m} \frac{\Phi (r) Y_{lm}(\theta, \phi )}{r} ~ e^{-i \omega t}.
\end{align}
By employing the tortoise coordinates, the perturbation equation becomes
\begin{align}
  &\Phi''(r_*) - (V_{\textrm{eff}} - \omega^2) \Phi(r_*) = 0, \; \;  \rmd r_* = \frac{1}{\sqrt{AB}} \rmd r , \label{eq:Veff} \\
  &V_{\textrm{eff}}(r) \!=\! \frac{l(l+1) A}{r^2} \!+\! \frac{1}{2\,r} \bigg(A' B + A B' \bigg)\! - \frac{1}{2}f_{\varphi_1 \varphi_1} A \, \mathcal{G} , \nonumber
\end{align}
where $l$ is the angular momentum. The system becomes unstable if the following condition is satisfied \cite{10.1119/1.17935, Myung:2019oua}
\begin{align}
  \int\limits^{\infty}_{r_h} dr \, \frac{1}{\sqrt{AB}} \, V_{\textrm{eff}}(r) < 0. \label{eq:stabilitycd}
\end{align}
This examines the condition that the system has negative energy under the scalar field perturbation. In this paper, we only investigate a sufficient condition for instability. When the condition below is satisfied, the Schwarzschild black hole (where $A = B = 1 - \frac{2M}{r}$ and $\phi = 0$) becomes unstable:
\begin{align}
  \alpha > \frac{5}{6} \big(2\, l(l+1) + 1 \big) M^2 = \alpha_\textrm{Sch.}
\end{align}
When $M=\frac{1}{2}$ and $l=0$, the critical value of $\alpha$ is $\alpha_\textrm{Sch.} =\frac{5}{24} \approx 0.2083$. A more rigorous examination of dynamical stability is studied in \cite{park2024}.

\section{Hairy black holes with global $U(1)$ symmetry}

For SSB to occur in this theory, three conditions must be met: (i) the coupling function $f(\varphi)$ must exhibit a certain symmetry; (ii) there must exist different ``global" minima $\dot{f}(\varphi_{\textrm{vac.}}) = 0$ depending on the coupling constant; and (iii) $\varphi_h$, the near-horizon value of $\varphi$, should lie near the vacuum of the potential: $\varphi_h \approx \varphi_{\textrm{vac.}} + \delta \varphi$.  
Our coupling function in (\ref{eq:f}) satisfies (i) and (ii), and we impose the boundary condition for $\varphi_h$ in accordance with (iii).

\subsection{Symmetric Phase}

When $\alpha$ is negative, the interaction potential $V$ has a minimum at $\varphi = \varphi^* = 0$, which we call a vacuum of the system. However, when $\alpha$ is positive, this point becomes a local maximum, and the system develops a new vacuum. Here, we search for hairy black hole solutions in the symmetric phase by assuming a small value of $\varphi$ (close to the vacuum expectation value) near the horizon and varying the value of $\alpha$ when $\lambda = \frac{1}{10}$.\\

\subsubsection{Equations of motion}

We write the equations of motion as follows:
\begin{align}
  & \frac{1}{2\, \kappa^2} \bigg(R_{\mu \nu} - \frac{1}{2}\, R \, g_{\mu \nu} \bigg) =  - \frac{1}{2} (\nabla_{\alpha} \varphi^* \nabla^{\alpha} \varphi)g_{\mu \nu} \nonumber \\
  & ~~ + \frac{1}{2}(\nabla_{\mu} \varphi^* \nabla_{\nu} \varphi + \nabla_{\mu} \varphi \nabla_{\nu} \varphi^*) \nonumber\\
  & ~~ - \frac{1}{2} (g_{\rho \mu} g_{\lambda \nu} + g_{\lambda \mu} g_{\rho \nu}) \, \eta^{\kappa \lambda \alpha \beta} \tilde{R}^{\rho \gamma}{}_{\alpha \beta} \nabla_{\gamma} \nabla_{\kappa} f, \label{eq:EE}\\
  & \nabla^{\alpha} \nabla_{\alpha} \varphi + \frac{\partial f}{\partial \varphi^*}  \mathcal{G} = 0, \; \; \nabla^{\alpha} \nabla_{\alpha} \varphi^{*} + \frac{\partial f}{\partial \varphi} \, \mathcal{G} = 0 . \label{eq:KG}
\end{align} 
Here $f=f(\varphi^*, \varphi)$ and $\tilde{R}^{\rho \gamma}{}_{\alpha \beta} = \eta^{\rho \gamma \sigma \tau} R_{\sigma \tau \alpha \beta} = \frac{\epsilon^{\rho \gamma \sigma \tau}}{\sqrt{-g}} R_{\sigma \tau \alpha \beta}$. Complex scalar fields are decomposed into two real scalar fields  
\begin{align}
  \varphi(r) = \frac{1}{\sqrt{2}} \big(\varphi_1(r) + i \, \varphi_2(r) \big), \label{eq:CtoR}
\end{align}
and the equations of motion (\ref{eq:EE})-(\ref{eq:KG}) are expressed in terms of $\varphi_1$ and $\varphi_2$. For the existence of a regular black hole, specific boundary conditions must be imposed near the horizon:
\begin{align}
  A(r) & \sim A_h \epsilon + \mathcal{O}(\epsilon^2) , \qquad B(r) \sim B_h \epsilon  + \mathcal{O}(\epsilon^2), \nonumber \\
  \varphi_i(r) & \sim \varphi_{ih} + {\varphi_{ih,1}} \epsilon  + \mathcal{O}(\epsilon^2) ,\label{eq:NHexp}
\end{align}
where $\epsilon = r - r_h$ is the expansion parameter, and $A_h, B_h, \varphi_{i h}$ and $\varphi_{i h,1}$ ($i=1,2$) are constants. We also set $\kappa^2 = 1/2$ hereafter. The following relations between the constants are obtained from the equations of motion:
\begin{align}
 &B_h = \frac{2} {r_h x_1}\big(1\pm \sqrt{1- x_1}\big) \bigg|_{r_h}, \nonumber\\
 &\varphi_{ih,1} = -\frac{r_h \varphi_{ih}}{4 x_2} \big(1\pm\sqrt{1- x_1}\big)\bigg|_{r_h} , \label{eq:NHcoeff}
\end{align}
where
\begin{align}
x_1 = \frac{96}{r_h^4} ( (f_{\varphi_1})^2 + (f_{\varphi_2})^2 ), \; \; x_2 = \varphi_1 f_{\varphi_1} +  \varphi_2 f_{\varphi_2} .
\end{align}
We found that the numerical solutions can only be generated for the minus sign before the root in $B_h$ and ${\varphi_{i h,1}}$. In addition, $\varphi''(r_h)$ diverges when the expression under the square root is zero. We exclude this case imposing the following condition:
\begin{align}
(1-x_1) |_{r_h} > 0 \label{eq:constraint2} .
\end{align}
Asymptotic flatness is required at infinity, which gives the following expansion:
\begin{align}
  &A \sim 1+\frac{A_1}{r} -\frac{A_1 \,x_3}{24\, r^3} + \cdots, \; \; B \sim 1+\frac{A_1}{r}+\frac{x_3}{4 \, r^2} + \cdots ,  \nonumber\\
  &\varphi_i \sim \varphi_{i\infty} + \frac{\varphi_{i,1}}{r} -\frac{A_1 \, \varphi_{i,2}}{2\, r^2}  + \cdots  \label{eq:vpexpInf} ,
\end{align}
where $x_3=(\varphi_{1,1})^2+ (\varphi_{2, 1})^2$, and all coefficients are constants. We identify the coefficient $A_1$ as the ADM mass of black holes such that $A_1 = - 2M$, and $\varphi_{i,1}$ is the scalar charge $\varphi_{i,1}=Q_i$. 

In the presence of symmetry, the conserved current is defined as
\begin{align}
  \partial_{\alpha} J^{\alpha} =  0, \qquad J_{\alpha} = i\, g\, \big(\varphi^* \, \partial_{\alpha} \varphi - \varphi \, \partial_{\alpha} \varphi^* \big).
\end{align}
The flux for a timelike hypersurface near the horizon reads
\begin{align}
  & \int_{\Sigma} J_{\alpha} n^{\alpha} \sqrt{-h}\, \rmd^3 y  = \int_{\Sigma} \bigg[g (\varphi_2 \partial_r \varphi_1 - \varphi_1 \partial_r  \varphi_2) \bigg] \nonumber \\
  & \qquad \times\bigg[\sqrt{A(r) B(r)} \, r^2 \sin \theta \, \rmd \theta \, \rmd \phi \, \rmd t \bigg] = 0 \label{eq:flux} ,
\end{align}
where $n_{\alpha}$ is a spacelike normal vector defining the timelike hypersurface $\Sigma$ with the induced metric $h$. As we assume that all solutions are regular with the expansion (\ref{eq:NHexp}) - (\ref{eq:constraint2}), the flux vanishes near the horizon. 

\subsubsection{Numerical Solutions}

\begin{figure}
\includegraphics[scale=0.36]{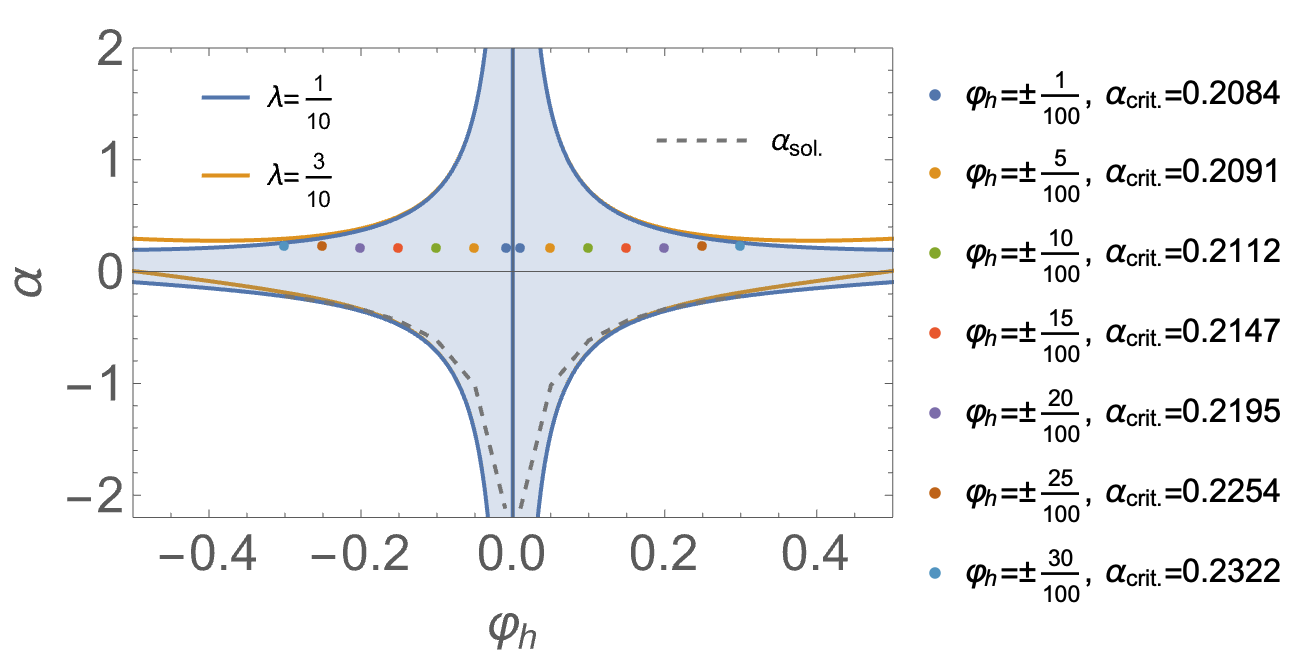} 
	\caption{The parameter space for $\alpha$ and $\varphi_h$ for $\lambda=\frac{1}{10}$ and $\lambda=\frac{3}{10}$ is shown in blue and orange, respectively.}
	\label{fig:SMMphase}
\end{figure}

The boundary conditions near the horizon are described by two coefficients, $A_h$ and $\varphi_{2h}=\varphi_{1h}=\varphi_h$ in (\ref{eq:NHexp})- (\ref{eq:constraint2}).  The value of $A_h$ is determined by the asymptotic flatness (\ref{eq:vpexpInf}), while $\varphi_h$ is constrained by (\ref{eq:constraint2}). This condition restricts the valid range of $\alpha$, $\lambda$, and $\varphi_h$ for a given $r_h$. The parameter space between these variables for the fixed values of $\lambda$ is drawn in Fig.\ref{fig:SMMphase}. In the search of the parameter space, we had difficulties to find the numerical solutions just above the lower solid lines. The gray dashed line denoted by $\alpha_{\textrm{sol.}}$ represents the first solution we found by gradually increasing the $\alpha$ value from the lower solid line. In contrast, numerical solutions are readily obtainable within the range $\alpha_{\textrm{sol.}} \leq \alpha < \alpha_{\textrm{max}}$, where $\alpha_{\textrm{max}}$ is the value corresponding to the upper solid line.
The sufficient condition for instability is given by Eq. (\ref{eq:stabilitycd}) and the critical value of $\alpha$ for the onset of instability is marked by dots labeled as $\alpha_{\textrm{crit.}}$ in Fig.\ref{fig:SMMphase}. Hairy holes become unstable when $\alpha_{\textrm{crit.}} \leq \alpha < \alpha_{\textrm{max}}$. 

\begin{figure}[h!]
	\subfloat[The metric solution for $\alpha_{\textrm{sol.}}$]{\includegraphics[scale=0.28]{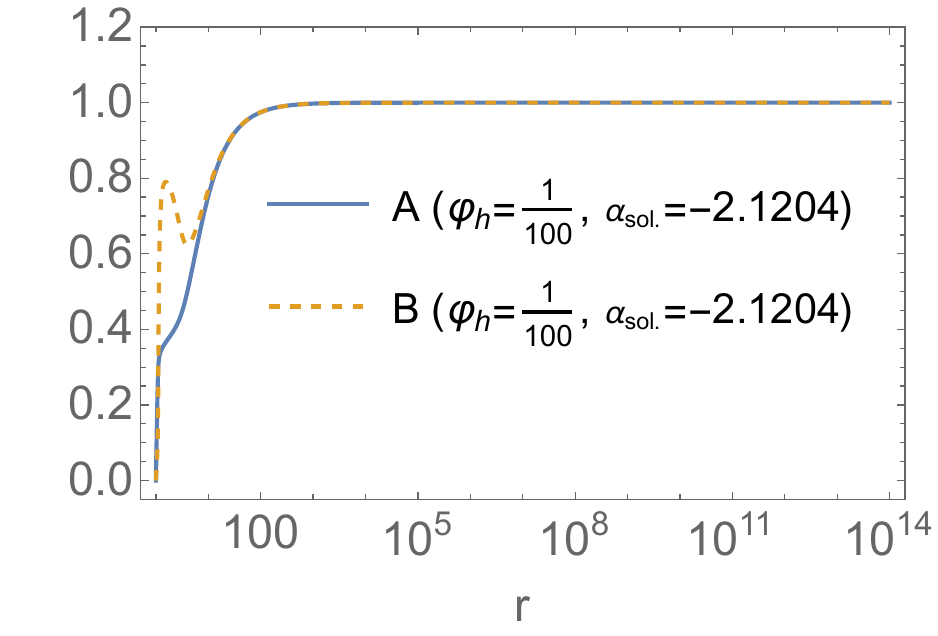}} 
	\subfloat[The metric solution for $\alpha_{\textrm{max.}}$]{\includegraphics[scale=0.28]{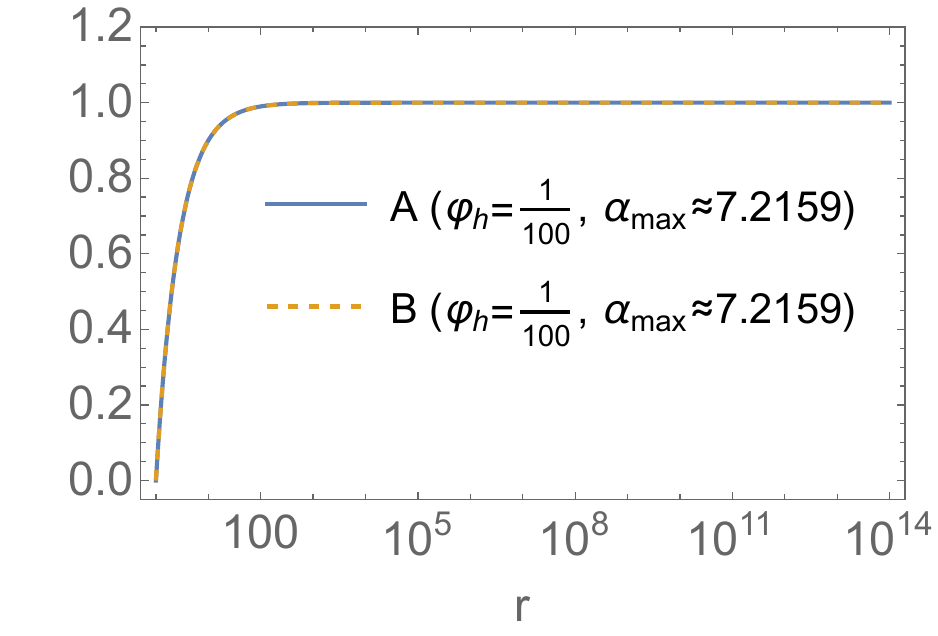}} 

	\subfloat[The scalar solution for $\varphi_h = \frac{1}{100}$]{\includegraphics[scale=0.28]{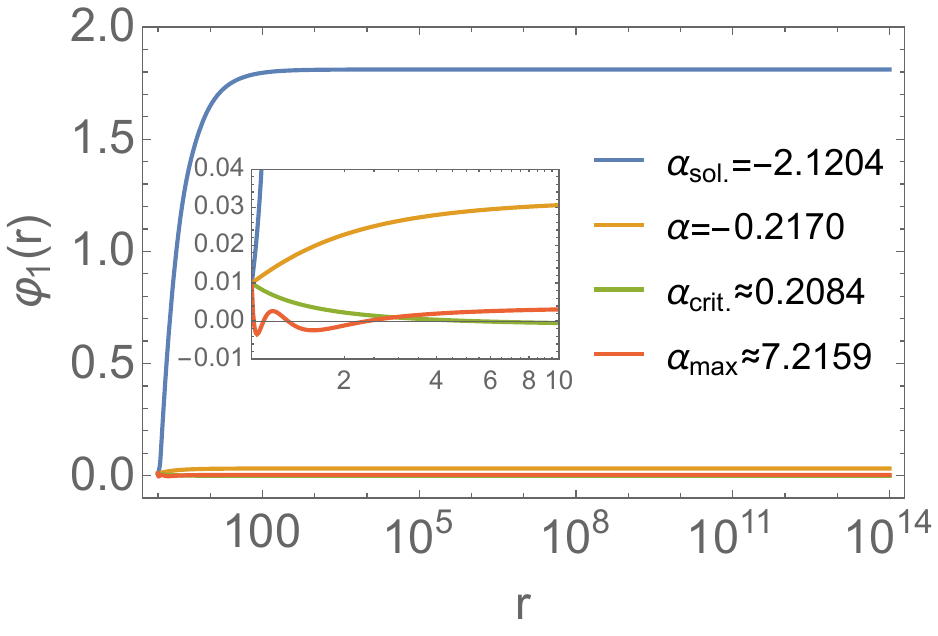}} 
	\subfloat[The effective potential, $V_{\textrm{eff.}}$]{\includegraphics[scale=0.28]{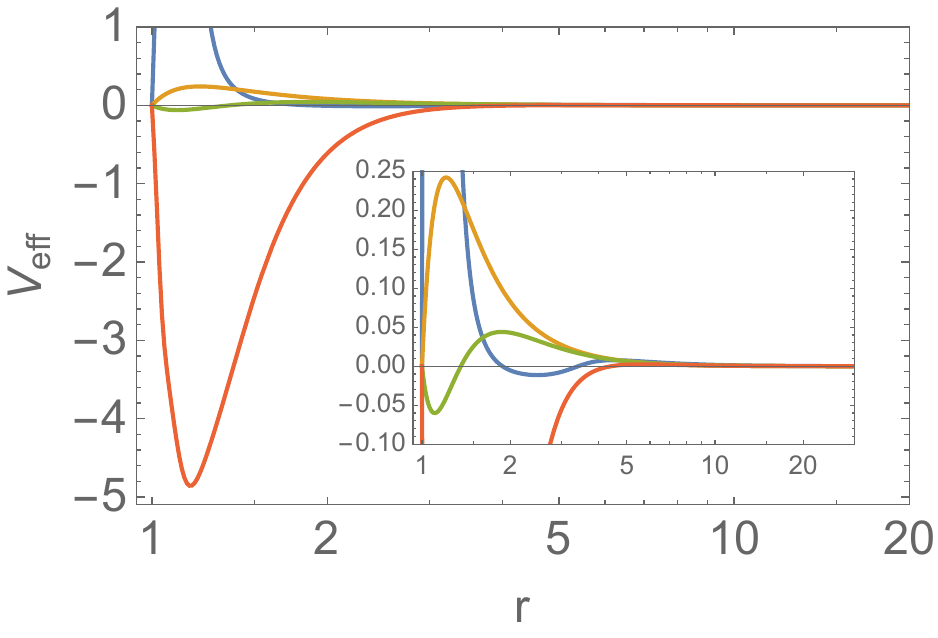}} 

	\subfloat[The black hole mass, $M$]{\includegraphics[scale=0.28]{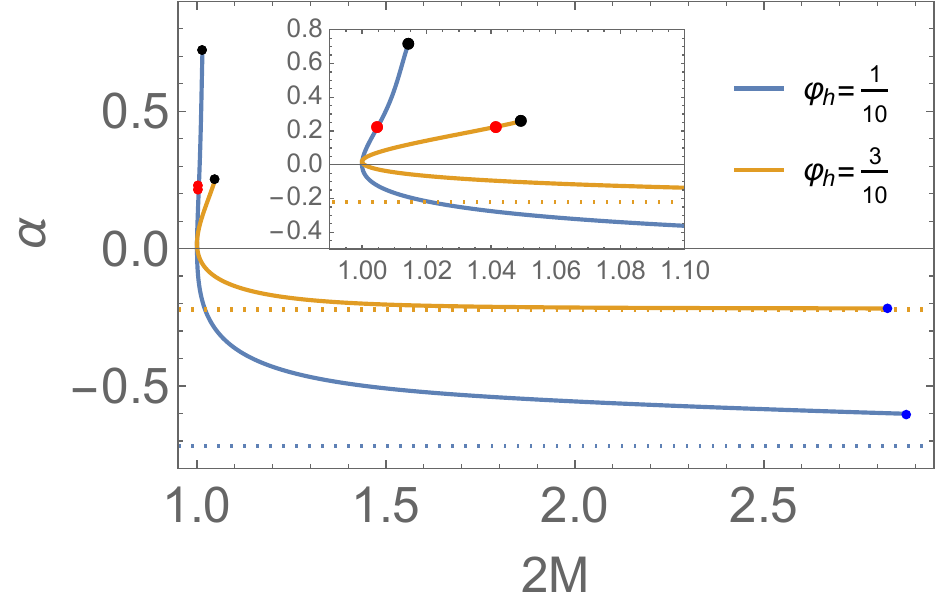}} 
	\subfloat[The scalar charge, $Q$]{\includegraphics[scale=0.28]{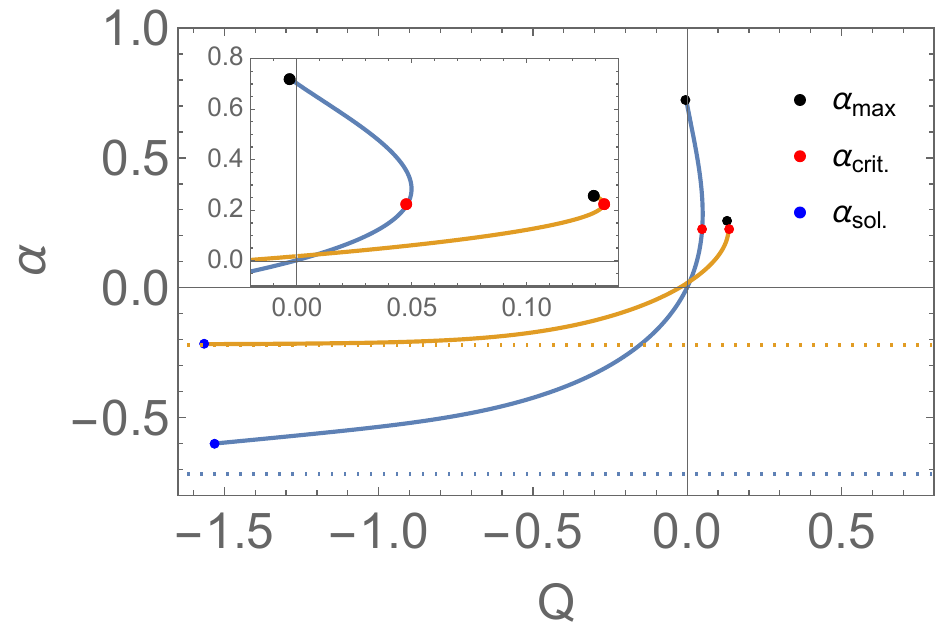}} 
\caption{Hairy black holes in the symmetric phase}
	\label{fig:SMM}
\end{figure}

The metric solutions are displayed in Fig.\ref{fig:SMM}(a) and (b). The metric with $\alpha_{\textrm{sol.}}$  displays severe non-monotonic behaviour near the horizon. Fig.\ref{fig:SMM}(c) shows scalar field solutions and Fig.\ref{fig:SMM}(d) illustrates the effective potential (\ref{eq:Veff}). As $\alpha$ approaches $\alpha_{\textrm{max}}$, the positive peak at $\alpha_{\textrm{sol.}}$ diminishes and eventually turns negative, signaling the system's instability. We calculated the mass ($M$) and scalar charge ($Q$) in Fig.{\ref{fig:SMM}} (e) and (f). The coloured dotted lines represent the minimum values of $\alpha$ $(\alpha_{\textrm{min.}})$ for the corresponding coloured solid lines. These figures show that the hairy black holes are always heavier than Schwarzschild black holes of the same radius, and the scalar charge becomes larger as $\alpha$ approaches $\alpha_{\textrm{sol.}}$. This might suggest that a large scalar charge causes the metric near the horizon to exhibit non-monotonic behaviour and the difficulties of finding hairy black holes below $\alpha_{\textrm{sol.}}$. Our findings might implicitly indicate the existence of a  relationship between mass and scalar charge similar to the mass-charge condition ($M \geq Q_e$) that prevents naked singularities in Reissner-Nordstr\"{o}m black holes.

Our numerical solutions were obtained using the ``NDSolve" in Mathematica with the options:  WorkingPrecision $\to 40$, PrecisionGoal $\to 40$, AccuracyGoal $\to 30$, MaxSteps $\to \infty$, and InterpolationOrder $\to$ All. These ensured high precision and accuracy in the numerical solutions. The numerical integration commenced from $r_h + \epsilon$, where $\epsilon=10^{-10}$, to $r=10^{14}$. The radial variables were rescaled by $r_h$, which was set to 1.

\subsection{Symmetry-Broken Phase}

When $\alpha$ is positive, the potential $V$ in (\ref{eq:action2}) has degenerate vacua, and the stable minima are determined by
\begin{align}
  \langle \varphi \rangle = v \, e^{i \beta},  \qquad v= \sqrt{\frac{\alpha}{2\, \lambda}} ,
\end{align}
where the vacuum states $ \langle \varphi \rangle$ are labelled by $\beta$. These ground states do not respect the symmetry of the Lagrangian, indicating that the symmetry is spontaneously broken. We expand the field around a ground state $v$ by reparameterizing it as follows
\begin{align}
  \varphi(r) = \bigg(v + \frac{\sigma(r)}{\sqrt{2}} \bigg)e^{i \theta(r)} .
\end{align}
Here, $\sigma(r)$ and $\theta(r)$ are physical fields because they describe excitations above the vacuum. In terms of new variables, the Lagrangian becomes
\begin{align}
  \mathcal{L}_{\varphi} = & - \frac{1}{2}\nabla_{\alpha} \sigma(r) \nabla^{\alpha} \sigma(r) \nonumber \\
  & - \bigg(v + \frac{\sigma(r)}{\sqrt{2}} \bigg)^2 \nabla_{\alpha} \theta(r) \nabla^{\alpha} \theta(r) + f(\sigma) \, \mathcal{G} ,
\end{align}
where
\begin{align}
  f(\sigma) = -\alpha \, \sigma (r)^2- \sqrt{\alpha\,  \lambda} \; \sigma (r)^3-\frac{\lambda}{4}\, \sigma (r)^4.
\end{align}
The system originally had one complex scalar field but now consists of one massive real scalar field $\sigma(r)$ and one massless real scalar field $\theta(r)$, the Goldstone boson. \\

\subsubsection{Equations of motion}

The equations of motion are given as
\begin{align}
  & \frac{1}{\kappa^2} \bigg[R_{\mu \nu} - \frac{1}{2} R g_{\mu \nu} \bigg] \!\!=\!\! \bigg[ \! -\frac{1}{2} (\nabla \sigma)^2 \!-\! \bigg( \! v + \frac{\sigma}{\sqrt{2}} \bigg)^2 \! (\nabla \theta)^2 \bigg]g_{\mu \nu} \nonumber\\
  & {~~~} + \nabla_{\mu} \sigma \nabla_{\nu} \sigma + 2 \bigg(v + \frac{\sigma}{\sqrt{2}} \bigg)^2 \nabla_{\mu} \theta \, \nabla_{\nu} \theta \nonumber \\
  & {~~~} - (g_{\rho \mu} g_{\lambda \nu} + g_{\lambda \mu} g_{\rho \nu}) \eta^{\kappa \lambda \alpha \beta} \tilde{R}^{\rho \gamma}{}_{\alpha \beta} \nabla_{\gamma} \nabla_{\kappa} f(\sigma), \\
  & \nabla^{\alpha} \nabla_{\alpha} \sigma - \sqrt{2} \bigg(v + \frac{\sigma}{\sqrt{2}}\bigg) \nabla^{\alpha} \theta\, \nabla_{\alpha} \theta + f_{\sigma}\, \mathcal{G} = 0, \\
  & \bigg(v + \frac{\sigma}{\sqrt{2}} \bigg)  \nabla^{\alpha} \nabla_{\alpha} \theta + \sqrt{2}\, \nabla^{\alpha} \sigma \nabla_{\alpha} \theta = 0.
\end{align}
The field $\theta(r)$ is decoupled from the system, and the solution for $\theta'(r)$ reads
\begin{align}
  \theta'(r) = \frac{c_2}{4 r^2 \sqrt{A(r)B(r)}} \bigg(v+ \frac{\sigma(r)}{\sqrt{2}}\bigg)^{-2},
\end{align}
where $c_2$ is an integration constant. If $c_2$ is not equal to zero, the solution is regular at infinity but singular at the horizon. If $c_2$ equals zero, the Goldstone boson becomes trivial. The corresponding flux for a timelike hypersurface reads
\begin{align}
  & \int\limits_{\Sigma} J_{\alpha} n^{\alpha} \sqrt{-h}\, \rmd^3 y = \int\limits_{\Sigma} \bigg[- 2\, g \bigg(v + \frac{\sigma(r)}{\sqrt{2}} \bigg)^2  \theta'(r) \bigg] \nonumber \\
  & ~~ \times \bigg[\sqrt{A(r) B(r)} \, r^2 \sin \theta \, \rmd \theta \, \rmd \phi \, \rmd t \bigg] = - \, 8 \pi \, g \, c_2.
\end{align}
Since the flux is zero, as given by (\ref{eq:flux}), $c_2$ shall also be zero. Thus, the singular Goldstone boson does not contribute to the scalar hairs. The hairy black holes in this theory can only have trivial Goldstone boson hairs. Consequently, the global U(1) symmetry is effectively equivalent to $Z_2$ symmetry in this perspective. 

We also impose the same boundary condition as in (\ref{eq:NHexp}) -(\ref{eq:NHcoeff}) near the horizon, but $\varphi_i (r)$ is replaced by $\sigma(r)$. The expansion coefficients in (\ref{eq:NHcoeff}) expand as
\begin{align}
x_1 = \frac{96}{r_h^4} (f_{\sigma})^2, \qquad x_2 = \sigma_h f_{\sigma} \label{eq:SMBxi} .
\end{align}
The boundary conditions are described only by the two independent variables $A_h$ and $\sigma_h$ and the coupling constants $\alpha$ and $\lambda$. The regularity condition (\ref{eq:constraint2}) is also required. 
The asymptotic flatness is imposed at infinity, and the metric functions and the scalar field are expanded in the same way as in (\ref{eq:vpexpInf}), replaced by $\varphi_{1,1}=\sigma_1$ and $\varphi_{2,1} = 0$. \\

\subsubsection{Numerical Solutions}

\begin{figure}[t!]
	\begin{center}
\includegraphics[scale=0.45]{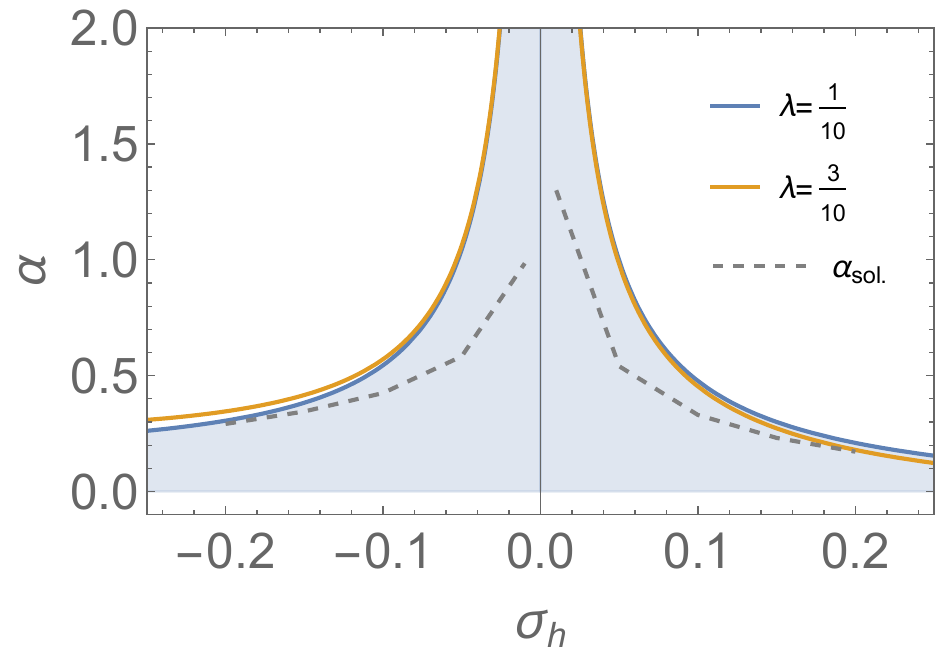}
	\caption{Parameter space for $\alpha$ and $\sigma_h$ for $\lambda=\frac{1}{10}$ and $\lambda=\frac{3}{10}$ is shown in blue and orange, respectively. }
	\label{fig:SMBphase}
	\end{center}
\end{figure}

We plot the parameter space satisfying (\ref{eq:constraint2}) with (\ref{eq:SMBxi}) for given $\lambda$ values in Fig.\ref{fig:SMBphase}. The parameter space is symmetric under $\varphi_h \rightarrow - \varphi_h$ in the symmetric phase. However, this symmetry is not present in the vaccua of this phase. For a given value of $\sigma_h$, the feasible range for $\alpha$ is $0 < \alpha < \alpha_{\text{max}}$, restricted by (\ref{eq:constraint2}).  Similar to the symmetric phase, our search for solutions to hairy black holes had difficulties to find beyond a certain value of $\alpha$, denoted $\alpha_{\textrm{sol.}}$, for given values of $\lambda$ and $\sigma_h$. In Fig. \ref{fig:SMBphase}, the gray dashed lines represent the last solutions we found by increasing values of $\alpha$ when $\lambda=\frac{1}{10}$.

\begin{figure}[b!]
		
	\subfloat[The metric sol. for small $\alpha$]{\includegraphics[scale=0.28]{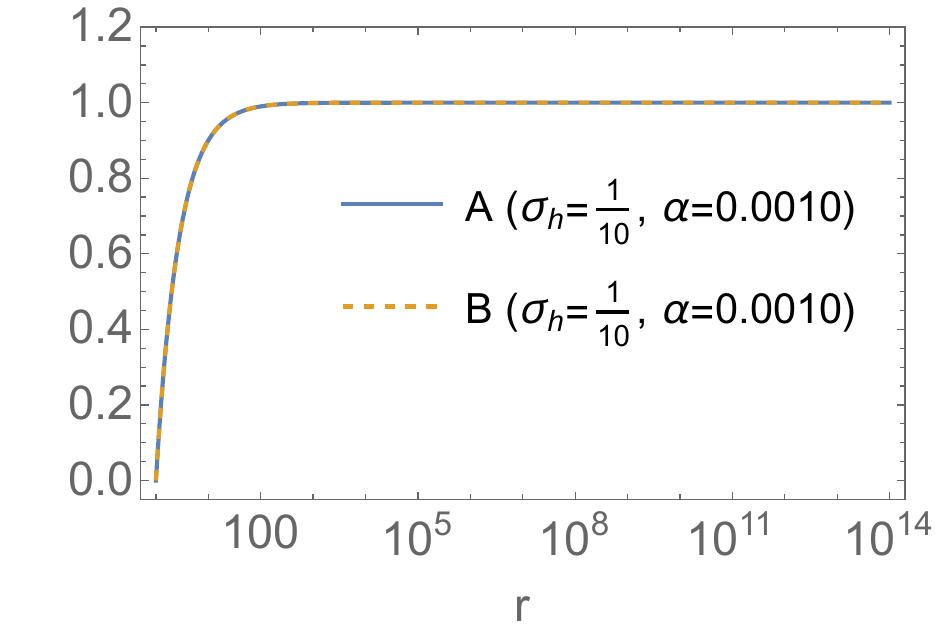}} \subfloat[The metric solution for $\alpha_{\textrm{sol.}}$]{\includegraphics[scale=0.28]{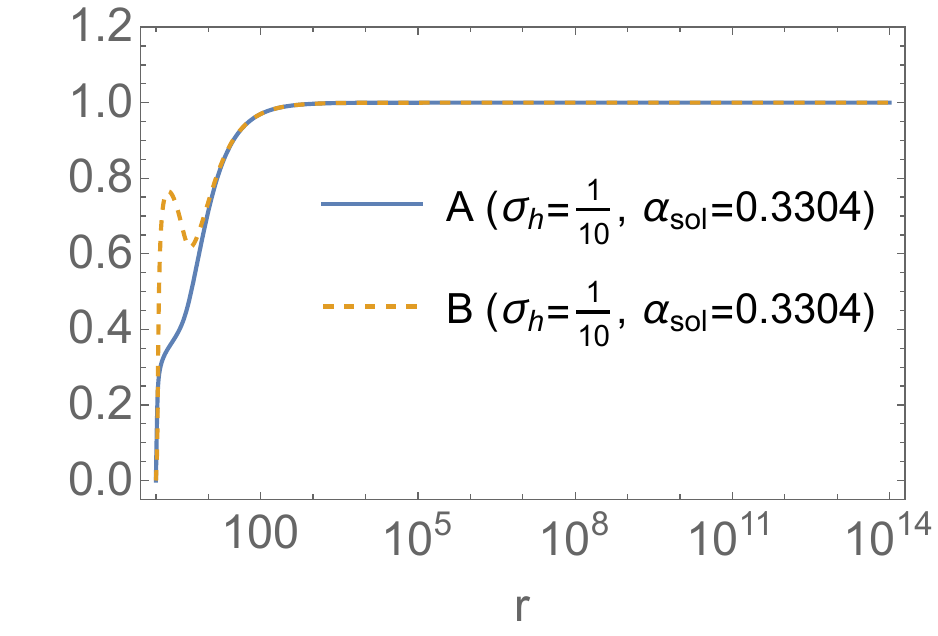}}

	\subfloat[The scalar field solution]{\includegraphics[scale=0.28]{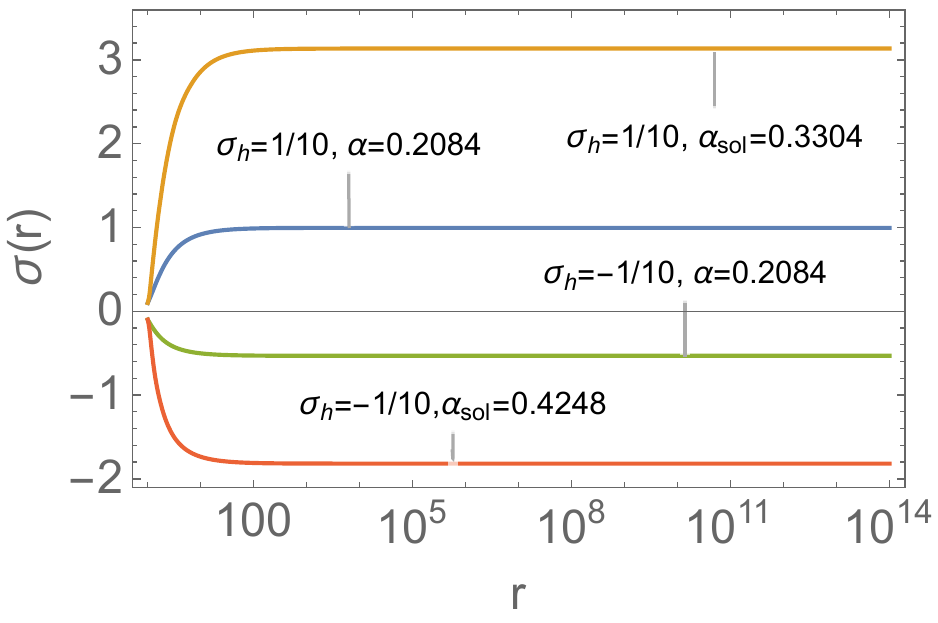}} 
	\subfloat[$V_{\textrm{effective}}$ for $\delta \sigma(r)$]{\includegraphics[scale=0.28]{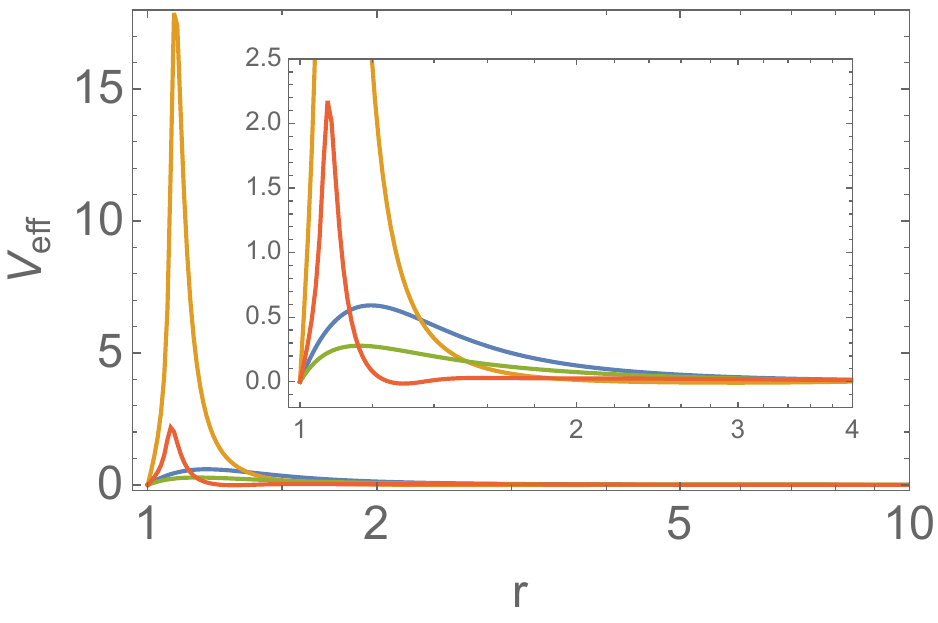}} 

	\subfloat[black hole mass]{\includegraphics[scale=0.28]{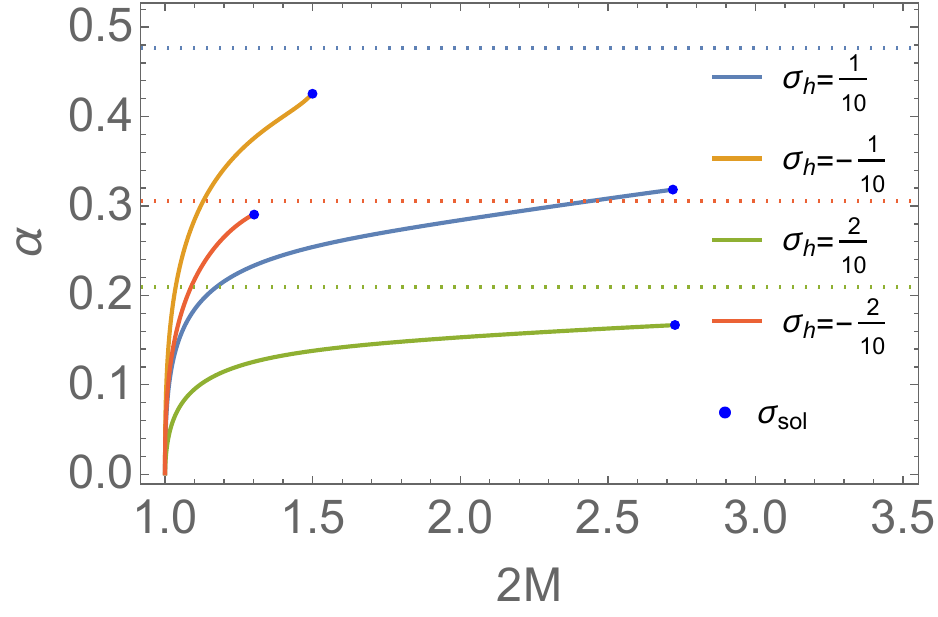}} 
	\subfloat[scalar charge]{\includegraphics[scale=0.28]{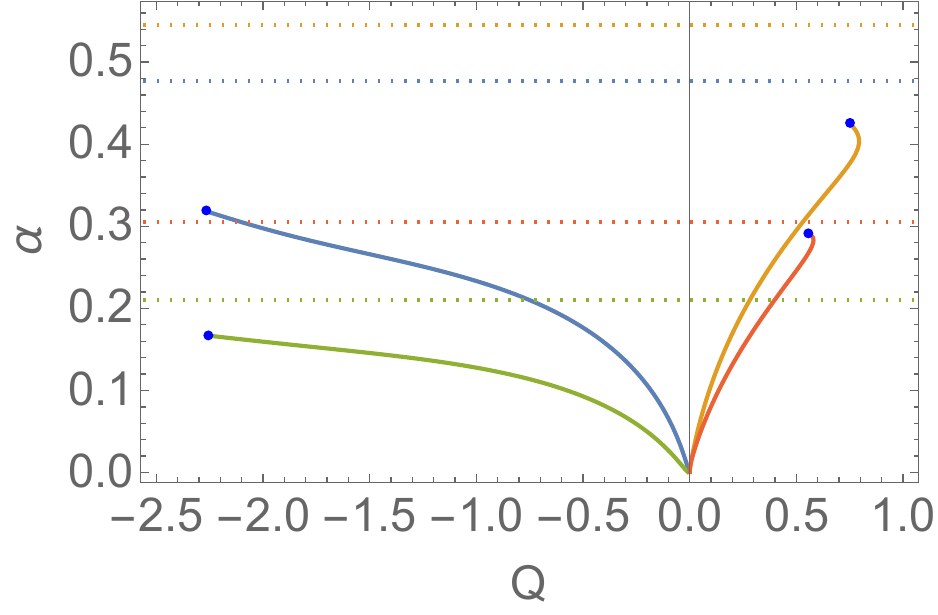}} 
\caption{Hairy black holes in the symmetry-broken phase}
	\label{fig:SMB}
\end{figure}

In Fig.\ref{fig:SMB}(a) and (b), we displayed the metric functions for $\sigma_h = \frac{1}{10}$. When $\alpha$ is small, the metric functions exhibit monotonicity, preserving the property of $g_{tt} g_{rr} \approx -1$. However, as $\alpha$ increases, these properties are lost near the horizon. The scalar fields in Fig.{\ref{fig:SMB}}(c), on the other hand, remain monotonic for all values of $\alpha \; (0 < \alpha \leq \alpha_{\textrm{sol.}})$ and their amplitude increases as $\alpha$ increases. Fig.\ref{fig:SMB}(d) displays the effective potentials in (\ref{eq:Veff}) with the solution presented in Fig.\ref{fig:SMB}(c). The effective potentials are always positive, indicating no direct indication of instability in the system. Given the same horizon radius ($r_h=1$), a hairy black hole always has a larger mass than a Schwarzschild black hole, as illustrated in Fig.\ref{fig:SMB}(e). The coloured dotted lines share the same values of $\sigma_h$ with the corresponding coloured solid lines and represent the maximum values of $\alpha$ from (\ref{eq:constraint2}). Fig.\ref{fig:SMB}(f) plots the scalar charge of a hairy black hole as a function of $\alpha$.\\

\section{Hairy black holes with local U(1) symmetry}

\begin{figure}[t!]
	\begin{center}
	\subfloat[The metric solution.]{\includegraphics[scale=0.28]{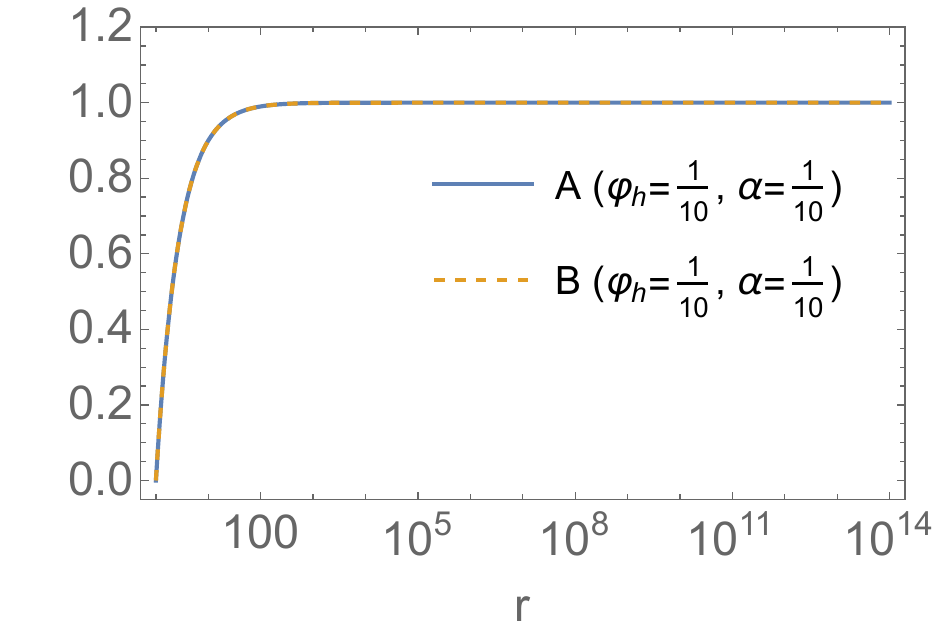}} \subfloat[The scalar and gauge field.]{\includegraphics[scale=0.28]{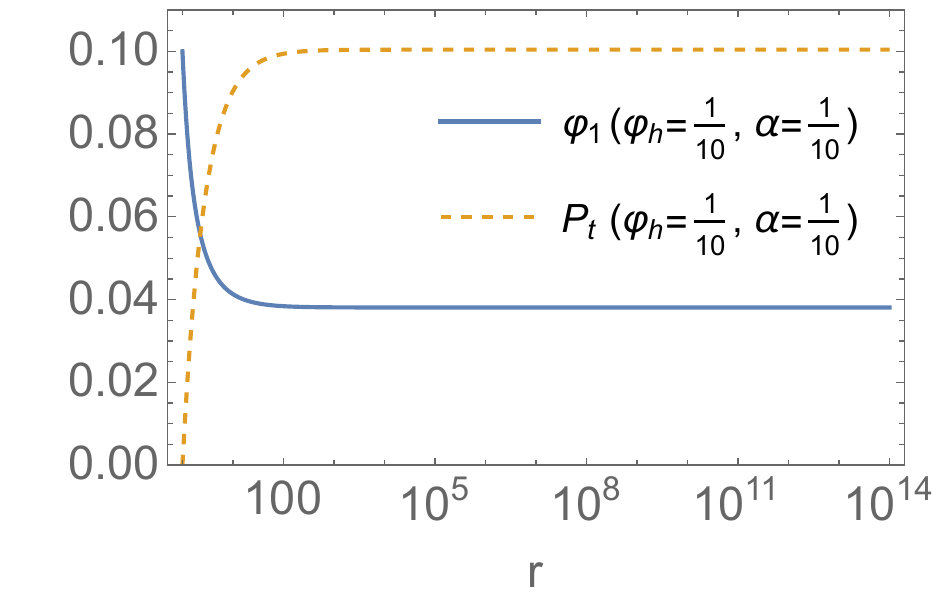}}
	\caption{numerical solutions for $q=0$}
	\label{fig:LU1sols}
	\end{center}
\end{figure}

Now, we examine the theory with a local U(1) symmetry by adding the gauge field to the Lagrangian:
\begin{align}
S= \int \rmd^4 x \sqrt{-g} \bigg[\frac{R}{2\, \kappa^2}  - \frac{F^2}{4}  - (D_{\alpha} \varphi)^{*} D^{\alpha} \varphi  + f(\varphi^*, \varphi) \mathcal{G} \bigg],
\end{align}
where $F_{\alpha \beta}=\nabla_{\alpha} P_{\beta} - \nabla_{\beta} P_{\alpha} $ ($P_{\alpha}$: an electromagnetic potential), $D_{\alpha} = \nabla_{\alpha}- i q P_{\alpha}$, and $q$ is an electric charge of scalar fields. This action is invariant under gauge transformations 
\begin{align}
\varphi(r) \rightarrow \varphi(r)\, e^{i \chi(r)}
\end{align} 
where $\chi(r)$ is an arbitrary function. Equations of motion include the gauge terms on the right side of (\ref{eq:EE})
\begin{align}
  & \frac{1}{2\, \kappa^2} \bigg(R_{\mu \nu} - \frac{1}{2}\, R \, g_{\mu \nu} \bigg) = \frac{1}{2} F_{\mu \delta} F_{\nu}{}^{\delta} - \frac{1}{8} F_{\alpha \beta}F^{\alpha \beta} g_{\mu \nu}  \nonumber\\ 
  &- \frac{1}{2} ((D_{\alpha} \varphi)^* D^{\alpha} \varphi)g_{\mu \nu} + \frac{1}{2}((D_{\mu} \varphi)^* D_{\nu} \varphi + D_{\mu} \varphi (D_{\nu} \varphi)^*) \nonumber\\
  &- \frac{1}{2} (g_{\rho \mu} g_{\lambda \nu} + g_{\lambda \mu} g_{\rho \nu}) \, \eta^{\kappa \lambda \alpha \beta} \tilde{R}^{\rho \gamma}{}_{\alpha \beta} \nabla_{\gamma} \nabla_{\kappa} f, \label{eq:EEq}\\
  & D^{\alpha} D_{\alpha} \varphi + \frac{\partial f}{\partial \varphi^*} \, \mathcal{G} = 0, \qquad (D^{\alpha} D_{\alpha} \varphi)^{*} + \frac{\partial f}{\partial \varphi} \, \mathcal{G} = 0 \label{eq:KGq}
\end{align} 
where $\nabla_{\alpha}$ is changed to $D_{\alpha}$ in (\ref{eq:EE}) and (\ref{eq:KG}). This action also gives the gauge field equations
\begin{align}
\nabla_{\mu} F^{\mu \nu} - i\, q\, (\varphi^* D^{\nu} \varphi - \varphi (D^{\nu} \varphi)^*) = 0. \label{eq:U1MW}
\end{align}
We take the ansatz of the electric potential to be $P= P_t(r) \textrm{d} t$. The $r$-component of (\ref{eq:U1MW}) requires 
\begin{equation}
(\varphi^* \varphi' - \varphi \, {\varphi^*}') =0
\end{equation}
where $'$ is a derivative with respect to $r$. To satisfy this condition, we demand $\varphi=\varphi_1 = \varphi_2$. We follow the same procedure as before, requiring the near-horizon expansion (\ref{eq:NHexp})
with
\begin{align}
  & B_h =\frac{32 y^2 P_h^2 + Y }{64 y^2 \left(6 A_h+P_h^2\right)}, \; \; \; {\varphi_{h,1}} = \frac{32 y^2 P_h^2-Y}{32  A_h y}
\end{align}
where
\begin{align}
  & Y = 4\, A_h + P_h^2 - \sqrt{z}, \; \; \; \; y = \varphi_h (\alpha - 2 \lambda \varphi_h^2) \\
  &z= \left(4\, A_h + (32\, y + 1 ) P_h^2 \right){}^2 - 512\, y\, A_h \left(6\, A_h+P_h^2\right), \nonumber
\end{align}
and additionally demand the regularity condition for the gauge field near the horizon
\begin{align}
 P_t \sim P_h \epsilon + \mathcal{O}(\epsilon^2). \label{eq:U1NHexp}  
\end{align}
Assuming asymptotic flatness and neutral scalar fields ($q=0$), the metric, gauge, and scalar fields expand as:
\begin{align}
A \sim & 1+\frac{A_1}{r}+\frac{P_1^2}{4\, r^2}-\frac{A_1 \varphi_1^2}{12 \, r^3} + \cdots, \\
B \sim & 1+\frac{A_1}{r}+\frac{P_1^2+2\, \varphi _1^2}{4 \, r^2} -\frac{A_1 \varphi _1^2}{4 \, r^3} + \cdots, \\
P \sim & P_{\infty }+\frac{P_1}{r}-\frac{P_1 \, \varphi _1^2}{12\, r^3}+ \cdots,\\
\varphi \sim & \varphi _{\infty } \!+\! \frac{\varphi _1}{r} \!-\! \frac{A_1 \varphi _1}{2 \, r^2} \!+\! \frac{\varphi _1 \left(4 \, A_1^2 - P_1^2 - \varphi _1^2\right)}{12\, r^3} + \cdots 
\end{align}
where all the coefficients are constants. This applies to hairy black holes with mass and electric charge but electrically neutral scalar hair ($q=0$). The corresponding numerical solutions are shown in Fig.\ref{fig:LU1sols}.

When $q \neq 0$, asymptotic expansions of the gauge and scalar fields give $P_{\infty} = P_1 = 0$ or $\varphi_{\infty} = \varphi_1 = 0$. This may mean that the gauge or scalar field decays faster than $1/r^n$ at infinity or that no electrically charged scalar hairy black hole solutions exist. Our numerical results suggest that the latter would be the case. We could not find any hairy black hole solutions with electrically charged scalar hair. Thus, this theory might not realize SSB associated with local U(1) symmetry.\\

\section{Conclusion}

Our work demonstrates that spontaneous symmetry breaking (SSB) can lead to the formation of "stable" hairy black holes from non-hairy ones in Einstein-scalar-Gauss-Bonnet (EsGB) theory with U(1) symmetry. The vanishing of the Gauss-Bonnet term at infinity confines the SSB process to the vicinity of the black hole horizon, where the interacting potential has a non-trivial effect. To support this argument, we generated hairy black holes irrespective of the sign of the coupling function in {\cite{Papageorgiou:2022umj}}, different from the spontaneous scalarization \cite{PhysRevLett.120.131104} which necessitates a positive coupling function for hairy black hole formation, and investigated the stability of hairy black holes for various values of $\alpha$ in the symmetric and symmetry-broken phase. 

We firstly discovered that when $2 M=1$ and $l=0$, Schwarzschild's black hole becomes unstable beyond $\alpha_{\textrm{Sch.}} = \frac{5}{24}$ and is expected to transform into hairy black holes. To describe this, we focus on the initial stage of forming hairy black holes, where a scalar field is beginning to develop. Thus small values for the scalar field are imposed near the black hole horizon, specifically $|\varphi_h| \leq \frac{3}{10}$ and $|\sigma_h| \leq \frac{3}{10}$ in the symmetric and symmetry-broken phase respectively.

For the global U(1) symmetry, hairy black holes become unstable if $\alpha_{\textrm{crit.}} \leq \alpha < \alpha_{\textrm{max.}}$. In the limit of a small field value of $\varphi_h$, the critical point of $\alpha$ at which the hairy black hole become unstable are approximately equal to the value for the Schwarzschild black hole ($\alpha_{\textrm{Sch.}} \approx \alpha_{\textrm{crit.}}$). Therefore, the Schwarzschild black hole might not evolve into hairy black holes in the symmetric phase. In the symmetry-broken phase, the effective potential takes positive values, indicating that the hairy black holes are all stable against the scalar field perturbation. Thus, we expect that the Schwarzschild black holes in the unstable range of $\alpha \; (\alpha > \alpha_{\textrm{Sch.}})$ would evolve into the hairy black holes in the symmetry-broken phase. 

For the local U(1), we found electrically charged black holes with neutral scalar hairs but no hairy black holes with electrically charged scalar hairs. This indicates that SSB associated with local U(1) is unlikely to occur within this theoretical framework.

Our approach applies to theories with diverse symmetries and admits generalization using a broader class of coupling functions. To advance our understanding, we must explore the possibility of identifying the scalar hair with one found in standard models. This connection could imply the existence of hairy black holes generated through spontaneous symmetry breaking.

\noindent\textbf{\textit{Acknowledgments.}}

We wish to thank Seong Chan Park, Chang Sub Shin, Sang Hui Im, Akihiro Ishibashi and Tae hyun Jung for the useful discussion. We were supported by the Institute for Basic Science (Grant No. IBS-R018-Y1). We appreciate APCTP for its hospitality during the completion of this work.

\bibliography{hairySSBbib}
\bibliographystyle{ieeetr}

\end{document}